\documentstyle[aps]{revtex}
\begin{document}

\twocolumn[\hsize\textwidth\columnwidth\hsize\csname
@twocolumnfalse\endcsname

\title{{\Large \bf A note on the foundation of relativistic
mechanics}\\[2mm]  I:  Relativistic observables and relativistic states}

\author{Carlo Rovelli \\
{\it Centre de Physique Th\'eorique, Luminy, F-13288 Marseille, EU}\\
{\it Physics Department, Pittsburgh
University, PA-15260, USA}}
\date{\today}
\maketitle    
\begin{abstract}
    Is there a version of the notions of ``state" and ``observable"
    wide enough to apply naturally and in a covariant manner to
    relativistic systems? I discuss here a tentative answer. 
\end{abstract} 
\vskip1cm]

\section{Introduction}    

The formalism of mechanics and its basic notions, such as ``state" and
``observable", are usually defined at first in the {\em
nonrelativistic\/} context.  One then modifies the formalism in order
to take gauge invariance into account \cite{dirac}.  Then one deals
with reparametrization invariant theories, such as the ones describing
relativistic gravitational systems, as special cases of gauge systems
in which the gauge includes spacetime transformations.  This way of
proceeding leads to a number of confusing issues and conceptual
difficulties concerning the notions of observable and state, in
particular in the gravitational quantum domain.  I think that these
difficulties are a consequence of the fact that the notions of state
and observable defined in a nonrelativistic manner at fixed external
time are being stretched and applied forcefully in the relativistic
domain, where they do not belong.

Here, I suggest a different approach.  I start from scratch,
considering definitions of ``state" and ``observable" given in a
naturally relativistic manner.  These relativistic notions are
introduced in Section \ref{pend}, taking a simple pendulum as an example. 
These notions are slightly, but significatively distinct from the
corresponding conventional nonrelativistic notions.  The main
difference is essentially that they do not refer to a moment of time. 
I think that in a relativistic context these relativistic notions of
state and observable are more natural to use than the usual ones,
which are intrinsically noncovariant.  The importance of the
relativistic notion of state has been stressed in particular by Dirac
\cite{Dirac2} and Souriau \cite{Souriau}, and certainly by many
others.  The relativistic notion of (``partial") observable has been
recently discussed for instance in \cite{r}.

It is a remarkable fact that the structure of mechanics simplifies
drastically when written in terms of the relativistic notions of
states and observable.  Hamiltonian mechanics takes a particularly
elegant and simple form.  This general and simple form 
was noticed and developed in different degrees
and with different motivations by many people, starting from Lagrange
\cite{lagrange} and including Arnold, Dirac and Souriau. 
I present it here in Section \ref{hamsyst} in an elementary form, based
on the relativistic notions of state and observable. 

In Section \ref{QM}, I present some remarks on general covariant
quantum mechanics, although the subject will be treated more in detail
elsewhere \cite{rr}.  In the companion paper \cite{c2}, I discuss the
applications of these ideas to field theory, and in particular in
general relativity.  I expect the relativistic definition of
observables and states discussed here to be especially relevant in
quantum gravity, where the conventional notions of position and time
are not available.

\section{Partial observables and Heisenberg states}\label{pend}

Imagine we want to study empirically the small oscillations of a
pendulum.  To this aim, we need two measuring devices.  A clock and a
device that reads the angle of elongation of the pendulum.  Let $t$ be
the reading of the clock and $\alpha$ the reading of the device
measuring the pendulum elongation.  Call the variables $t$ and
$\alpha$ the {\em partial observables} \cite{r}, or, if there is no
ambiguity, simply the {\em observables} of the system.  Let $\cal C$
be the two-dimensional space with coordinates $t$ and $\alpha$.  Call
$\cal C$ the {\em extended} or {\em relativistic configuration space}
of the pendulum, or simply, if there is no ambiguity, the {\em
configuration space} of the pendulum.

We perform a sequence of measurements.  We get a sequence of pairs
$(t,\alpha)$.  We call each such pair a {\em correlation} between the
two partial observables $t$ and $\alpha$.  Each pair determines a
point in the extended configuration space $\cal C$.  The reason we can
do science is the fact that experience shows we can find mathematical
relations characterizing the sequences of correlations that are
physically realized.

These relations have the following form.  We perform a sequence of
measurements of $t$ and $\alpha$, and find that the points
representing the measured pairs sit on a curve in the space $\cal C$. 
Let this curve be represented as a relation in $\cal C$
\begin{equation}
    f(\alpha,t)=0. 
\label{uno}
\end{equation} 
Call this curve a {\em motion} of the system.  Thus a motion is a
certain relation between observables.  We then disturb the pendulum
(push it with a finger) and repeat the entire experiment over.  At
each repetition of the experiment, a different curve is found.  That
is, a different mathematical relation of the form (\ref{uno}) is
found.  However (this is the central point), experience shows that the
space of possible curves is limited: indeed, it is a two-dimensional
space: there is just a two-dimensional space of curves that are
realized in nature.  

Denote $\Gamma$ this two-dimensional space of curves, and let $A$ and
$\phi$ be coordinates on $\Gamma$.  The two-dimensional space $\Gamma$
with coordinates $A$ and $\phi$ is the {\em space of the motions},
or the {\em Heisenberg phase space\/}, or, if there is no ambiguity,
simply the {\em phase space\/} of the pendulum.  We denote a point in
$\Gamma$ as a {\em motion} of the pendulum, or a {\em Heisenberg
state}, or simply, if there is no ambiguity, a {\em state}. 

The equation that captures the experimental relations in full has
therefore the form
\begin{equation}
    f(\alpha,t;\ A,\phi)=0. 
    \label{eq:f}
\end{equation}
That is: for each point in $\Gamma$ with coordinates $(A,\phi)$, we
have a curve in $\cal C$.  Equation (\ref{eq:f}) is the {\em evolution
equation} of the system.

Concretely, in the case of the (small oscillations of a frictionless)
pendulum, the evolution equation is 
\begin{equation}
    f(\alpha,t;A,\phi)=\alpha - A\sin(\omega t+\phi) =0.  
    \label{eq:f2}
\end{equation}
The pair $(A,\phi)$ labels different sequences of measurements.  For
each sequence, $(A,\phi)$ determines a curve in the $(t,\alpha)$
plane, expressing the measured, or predicted, correlation between $t$
and $\alpha$.  Thus a state (a pair $(A,\phi)$) determines a motion of
the pendulum (a specific relation between $t$ and $\alpha$) via the
evolution equation (\ref{eq:f2}).  Each time we disturb the pendulum
by interacting with it, or each time we start a new experiment over
with the same pendulum, we have a new state.  On the other hand, the
state remains the same (disregarding quantum theory) if we just
observe the pendulum and the clock without disturbing them.

Summarizing: \emph{each state in the phase space determines a relation
between the observables in the configuration space}.  Each such
relation is called a motion.  The set of these relations is captured
by the evolution equation (\ref{eq:f}), namely by the vanishing of a
function 
\begin{equation}
    f: {\Gamma}\times{\cal C} \to R. 
\label{pnd}
\end{equation}
The evolution equation $f=0$ expresses all the predictions that can be
made using the theory.  Equivalently, these predictions are completely
captured by fixing the surface (\ref{pnd}) in the Cartesian product
${\Gamma}\times{\cal C} $ of the phase space with the configuration
space.

Concretely, predictions can be obtained as follows.  We first need to
perform enough measurements to deduce $A$ and $\phi$, namely to find
out the state.  Once the state is determined or guessed, the evolution
equation (\ref{eq:f2}) predicts the allowed correlation between the
observables $t$ and $\alpha$ in any subsequent measurement.  These
predictions are valid until the pendulum is disturbed.  

The $({\cal C}, \Gamma, f)$ structure described above for the example
of the pendulum is completely general, and is present in all
relativistic and nonrelativistic fundamental systems.  All
fundamental systems can be described (at the accuracy at which quantum
effects can be disregarded) by making use of these fundamental
concepts:
\begin{itemize} 
    \item[(i)] The \emph{configuration space} $\cal C$, of the
    \emph{observables} of the theory.  
    \item[(ii)] The \emph{phase
    space} $\Gamma$ of the \emph{states} of the theory.  
    \item[(iii)]
    The \emph{evolution equation} of the theory $f=0$, where  
    $ f: {\Gamma}\times{\cal C} \to V$.
\end{itemize}
$V$ is a (finite or infinite dimensional) vector space.  The state in
the phase space $\Gamma$ is fixed until the system is disturbed.  Each
state in $\Gamma$ determines (via $f=0$) a motion of the system,
namely a relation (or a set of relations if $dim(V)>1$), between the
observables in $\cal C$.  The task of mechanics is to find such a
description for all physical systems.

The construction of this description for a given system is
conventionally separated in two steps.  The first step, kinematics,
consists in the specification of the observables that characterize the
system.  Namely the specification of the configuration space $\cal C$
and its physical interpretation.  (Physical interpretation means the
association of certain coordinates on $\cal C$ with certain measuring
devices.)  The second step, dynamics, consists in finding the phase
space $\Gamma$ and the function $f$ that describe the correlations in
the system. 

Notice that the notions of instantaneous state, evolution in time,
or observable at fixed time, play no role in the general definitions of
observable and state given.  This is why these definitions apply 
naturally to a special and generally relativistic context.  Of course, the
usual notions can be easily recovered for nonrelativistic systems, as
will be discussed in Section \ref{recov}.

\section{relativistic hamiltonian systems}\label{hamsyst}

The relativistic notions of state and observable defined above find
their natural home in the relativistic, or presymplectic hamiltonian
formalism, which I summarize here.

Virtually all fundamental physical systems can be described by
hamiltonian mechanics (I suppose as a consequence of the fact that
they are the classical limit of a quantum system).  That is, it turns
out that once the kinematics is known, namely once $\cal C$ has been
determined, the dynamics ($\Gamma$ and $f$) is completely determined
by a function $H$ on the cotangent space $\Omega=T^*{\cal C}$.
\begin{equation}
      H: T^*{\cal C} \to  W. 
\end{equation}
where $W$ is a vector space.  (The generalization of this structure to
field theory is discussed in the companion paper \cite{c2}.)  The pair
$({\cal C},H)$ fully determines the mechanics of the physical system. 
The pair $({\cal C},H)$ is a {\em covariant dynamical system}.  The
function $H$ can be called as {\em the constraints}, or {\em the
relativistic hamiltonian}, or if there is no ambiguity simply as the
{\em hamiltonian}.  Remarkably, all fundamental systems in nature seem
to have precisely this structure.

There are several equivalent descriptions of the way in which
$H$ determines the phase space $\Gamma$ and $f$:

\subsubsection{From $({\cal C}, H)$ to $({\cal C}, \Gamma, f)$:
Hamilton equations} In coordinates, $({\cal C},H)$ determines the
evolution equations as follows.  Let $q^a$, with $a=1,\ldots, n$, be
coordinates on ${\cal C}$, $p_a$ be the corresponding momenta in
$\Omega=T^*{\cal C}$ and let $H^i(q^a,p_{a})$, with $i=1,\ldots,
m=dim(V)$, be the components of $H$.  The motions are given by the
$m$-dimensional surfaces in ${\cal C}$, coordinatized by $m$
parameters $\tau_{i}$, obtained by solving the Hamilton equations
\begin{eqnarray}
    {\partial q^a(\tau_{i})\over\partial\tau_{i}} 
    ={\partial H^i \over\partial p_a}, \ \ \ 
    {\partial p_a(\tau_{i})\over\partial\tau_{i}} 
     =-{\partial H^i \over\partial q^a}, \ \ \ 
    H^i=0.
    \label{eq:hem}
\end{eqnarray}

\subsubsection{ From $({\cal C}, H)$ to $({\cal C}, \Gamma, f)$: the
geometrical way} Every cotangent space carries the natural symplectic
form $d\theta$, where $\theta=p_{a}dx^a$ is the Poincar\'e one-form of
the cotangent bundle.  The equation $H=0$ defines a surface $\Sigma$
in $T^*{\cal C}$.  The restriction $\omega$ of $d\theta$ to this
surface is a degenerate two-form with null directions.  The integral
surfaces of these null directions are called the \emph{orbits} of
$\omega$ on $\Sigma$.  The phase space $\Gamma$ is defined as the
space of these orbits.  Each such orbit projects down from $T^*{\cal
C}$ to $\cal C$ to give a subspace of $\cal C$, namely a set of
relations on $\cal C$.  Namely a motion.  To compute the orbits, we
have to integrate the (multi- )vector field $X$ on $\Sigma$ which 
is in the kernel of $\omega$, namely which satisfies the equation
\begin{equation}
	   \omega(X)=0
    \label{eq:omegaX}
\end{equation}
Equation (\ref{eq:omegaX}) is an elegant geometrical way of writing
the Hamilton equations (\ref{eq:hem}).  Notice that the system is
completely defined just by the pair $(\Sigma, \omega)$: a space and a
form over it.  The form $\omega$ is closed and degenerate, or
presymplectic, hence the denomination presymplectic for this
formalism.  In the companion paper \cite{c2}, in particular, I will
discuss a very simple, compact and elegant formulation of general
relativity in terms of a $(\Sigma, \omega)$ pair.

\subsubsection{From $({\cal C}, H)$ to $({\cal C}, \Gamma, f)$:
via Hamilton-Jacobi} Consider the (generalized) Hamilton-Jacobi
system of $m$ partial differential equations on $\cal C$
\begin{equation}
      H\left(q^a,\frac{\partial  S(q^a)}{\partial q^a}\right)=0. 
\end{equation}
Let $S(q^a,Q^a)$ be a $n$-parameter family of (independent, in a
suitable sense) solutions.  Then pose
\begin{equation}
  f^n(q^a,Q^a,P_a)=\frac{\partial S(q^a,Q^a)}{\partial Q^a}-P_a. 
\label{fn}
\end{equation} 
In general, only $n-k$ of these equations are independent, and only
$2(n-k)$ of the constants $X^a,P_a$ are independent.  The constants
$Q^a,P_a$ coordinatize thus a $2(n-k)$ dimensional space $\Gamma$. 
This is the phase space, and (\ref{fn}) defines $f$.

\vskip.2cm

This ``relativistic", or ``presymplectic" formulation of mechanics is
elegant, very well operationally founded, and far more general
than the conventional nonrelativistic formulation based on the triple
$(\Gamma_{nr},\omega_{nr}, H_{0})$, where $\Gamma_{nr}=T^{*}Q$ is the
space of the initial data, $\omega_{nr}$ its symplectic form and
$H_{0}$ the nonrelativistic hamiltonian.  The relation between the two
formulations is discussed below.

\subsection{Nonrelativistic systems}

For a nonrelativistic systems, the extended configuration space has
the structure 
\begin{equation}
{\cal C}= {I\!\!R}\times Q,
    \label{eq:CCnr}
\end{equation}
and its coordinates $q^a=(t,q^i)$, with $i=1,\ldots n$--1, are the time
variable $t\in {I\!\!R}$ and the physical degrees of freedom variables
$q^i\in Q$, where $Q$ is the usual non relativistic configuration
space.  The cotangent space $\Omega=T^*{\cal C}$ has coordinates
$(q^a,p_{a})=(t,q^i,p_{t},p_{i})$, and
\begin{equation} 
    H= p_{t}+H_{0} 
    \label{eq:HH0}
\end{equation} 
where $H_{0}$ is the nonrelativistic hamiltonian.  The surface
$\Sigma$ turns out to be
\begin{equation}
    \Sigma= {I\!\!R} \times \Gamma_{nr} 
    \label{eq:SigmaGammanr}
\end{equation}
where the coordinate on ${I\!\!R}$ is the time $t$ and $\Gamma_{nr}=T^*Q$
is the usual phase space.  The restriction of $d\theta$ to this surface
has the form
\begin{equation}
    \omega= -dH_{0}\wedge dt +\omega_{nr}. 
    \label{eq:omegaomeganr}
\end{equation}
The evolution equations (\ref{eq:hem}) or (\ref{eq:omegaX}) define the
physical motions in parametrized form $q^a(\tau) =
(t(\tau),q^i(\tau))$.  Only the dependence of $q^i$ on $t$ determined
implicitly by these functions has physical significance: not the
explicit dependence of $t$ and $q^i$ on $\tau$.  That is, the physics
is contained in a curve in $\cal C$, not in the way this curve is
parametrized.

We can take the vector field $X$ with the form
\begin{equation}
    X={\partial\over\partial t}+X_{nr}
    \label{eq:XXnr}
\end{equation}
where $X_{nr}$ is a vector field on $\Gamma_{nr}$. In this case, 
equation (\ref{eq:omegaX}) reduces to  
the equation 
\begin{equation}
    \omega_{nr}(X_{nr})=-dH_{0},
    \label{eq:omegaXnr}
\end{equation}
which is the geometric form of the well known nonrelativistic form of 
the Hamilton equations 
\begin{eqnarray}
    {\partial q^a(t)\over\partial t} 
    ={\partial H_{0} \over\partial p_a}, \ \ \ \ \ 
    {\partial p_a(t)\over\partial t} 
     =-{\partial H_{0} \over\partial q^a}. 
    \label{eq:hamnr}
\end{eqnarray}
Therefore the nonrelativistic hamiltonian $H_{0}$ generates evolution
in the time $t$.  More precisely: $H$ determines how the variables in
$Q$ are correlated to the variable $t$.  This can be expressed as
``how the variables in $Q$ evolve in time".

\subsection{Discussion}
\label{recov}

The definition of state and observable given in Section \ref{pend}
differ from the one that commonly used in the nonrelativistic
context.  In the nonrelativistic context the special variable $t$ is
chosen to play a peculiar role.

The usual nonrelativistic definition of {\em state} refers to the
properties of a system {\em at a certain moment of time}.  Denote this
conventional notion of state as the ``instantaneous state".  The space
of the instantaneous states is the usual nonrelativistic phase space
$\Gamma_{nr}$.  For instance, we fix the the value $t=t_{0}$ of the
time variable, and characterize the instantaneous state in terms of
the initial data, for the pendulum position and momentum $(\alpha_{0},
p_0)$, at $t=t_{0}$.  Then $(\alpha_{0}, p_0)$ are coordinates on
$\Gamma_{nr}$. 

Instead, in Section \ref{pend} I have defined a relativistic state as
a solution of the equation of motion.  (This is true for a system
without gauge invariance.  If there is gauge invariance, a state is a
gauge equivalence class of solutions of the equations of motion). 
Thus, the relativistic phase space $\Gamma$ defined above is the space
of the solutions of the equations of motion.

Once a value $t_{0}$ of the time has been fixed, there is a one-to-one
correspondence between initial data and solutions of the equations of
motion.  Indeed, each solution of the equation of motion determines
initial data at $t_{0}$; and each choice of initial data at $t_{0}$
determines uniquely a solution of the equations of motion.  Therefore
there is a one-to-one correspondence between instantaneous states and
relativistic states.  (This one-to-one correspondence is also present
if there is gauge invariance, because a choice of initial data
determines uniquely a gauge equivalence class of solutions.)  The
conventional phase space of the initial data, $\Gamma_{nr}$, with
coordinates $(\alpha_{0}, p_{0})$ is therefore isomorphic with the
relativistic phase space $\Gamma$, with coordinates $(A, \phi)$.  The
two be can be identified after having chosen a value $t=t_{0}$ of the
time variable.  The identification map is given in the
example of the pendulum by
 \begin{eqnarray}
     &(A,\phi) & \longmapsto  (\alpha_{0}(A,\phi), p_{0}(A,\phi)) : \\ 
     &\alpha_{0}(A,\phi) &\ \ =\  A \sin(\omega t_{0}+\phi),
     \label{eq:a}  \\
     & p_0(A,\phi) &\ \  =\   \omega m A \cos(\omega t_{0}+\phi). 
     \label{eq:pa}  
 \end{eqnarray}
Thus $(\alpha_{0}, p_{0})$ can be simply seen as coordinates on
$\Gamma$, and mathematically the phase space $\Gamma$ defined above is
essentially the same space as the nonrelativistic phase space: $\Gamma
\sim \Gamma_{nr}$.  But the physical interpretation of the two is
quite different: a point of $\Gamma$ is not seen as representing the
instantaneous state of the pendulum, but rather as representing the
full history of the pendulum that evolves from that instantaneous
state.

An important remark is that in a nonrelativistic system the space of
the orbits $\Gamma$ is in one to one correspondence with the cotangent
space $\Gamma_{nr}$.  Therefore the cotangent space
$\Gamma_{nr}=T^{*}Q$ is the ``natural arena" for nonrelativistic
hamiltonian mechanics and also the space of the motions.  The phase
space plays therefore a double role in nonrelativistic hamiltonian
mechanics: as the arena of hamiltonian mechanics and as the space of
the states.  In the relativistic context, in general, this double role
is lost: one must distinguish the cotangent space over which
hamiltonian mechanics is defined ($\Omega=T^*{\cal C}$) from the space
of the space of the states $\Gamma$ (which we call phase space).  This
distinction becomes important in field theory, where the space on
which hamiltonian mechanics is formulated can be finite dimensional
even if the phase space is infinite dimensional \cite{c2}.

In a nonrelativistic system, $X_{nr}$ generates a one-parameter group
of transformation in $\Gamma$: this sends a state to a state in which
the observables are the same, except for $t$ which is changed.  This
transformation group is the hamiltonian flow of $H_{0}$ on $\Gamma$. 
Instead of having the observables in $Q$ depending on $t$, one can
shift perspective and view the observables in $Q$ as time independent
objects and the states in $\Gamma$ as time dependent objects.  This is
a classical analog of the shift from the Heisenberg to the
Shr\"odinger picture in quantum theory, and can be called the
``classical Schr\"odinger picture". 

On the other hand, in a general relativistic system there is no
special ``time" variable, ${\cal C}$ does not split naturally as
${\cal C}={I\!\!R}\times Q$, the constraints do not have the form
$H=p_{t}+H_{0}$ and the description of the correlations in terms of
``how the variables in $Q$ evolve in time" is not available in
general.  For these systems the classical Schr\"odinger picture, in
which states evolve in time, is not available: only the relativistic
notions of state and observable considered here (partial observables
and Heisenberg states) make sense.

For these systems, dynamics is not a theory of the evolution of
observable quantities in time.  It is the theory of the correlations
between partial observables.

\subsection{Simple examples}
\subsubsection{Pendulum}

On $T^*{\cal C}$ we can put canonical coordinates $(t,\alpha,
p_{t},p)$.  The function $H$ is (with mass=1)
\begin{equation}
   H(t,\alpha, p_{t},p_{\alpha}) = p_{t}+\frac{p^2
+\omega^2\alpha^2}{2}\equiv  p_{t}+H_{0}(\alpha, p). 
\end{equation}
Equations (\ref{eq:hem}) give 
\begin{eqnarray}
    \frac{d t(\tau)}{d\tau} =  1, \  \ 
    \frac{d \alpha(\tau)}{d \tau} =  p(\tau) ,\ \  
\frac{d p(\tau)}{d \tau} = -\omega^2 \alpha(\tau). 
\end{eqnarray} 
whose solutions project down to the $\cal C$ space to the trajectories
\begin{equation}
    f(\alpha,t;A,\phi)=\alpha - A\sin(\omega t+\phi) =0.  
\end{equation}
Thus the phase space $\Gamma$ has coordinates $A$ and $\phi$ and we
recover (\ref{eq:f2}) for $f$.

\subsubsection{Relativistic particle}\label{Lorentz}

The configuration space $\cal C$ is Minkowski space $\cal M$, with
coordinates $x^\mu$. The dynamics is given by the hamiltonian
$H=p^2-m^2$, which defines the mass-$m$ Lorentz hyperboloid ${\cal
K}_{m}$.  The constraint surface $\Sigma$ is therefore given by
$\Sigma={\cal M}\times {\cal K}_{m}$.  The null vectors of the
restriction of $d\theta=dp_{\mu}\wedge dx^\mu$ to $\Sigma$
are therefore 
\begin{equation}
X=p_{\mu}\frac{\partial}{\partial x^\mu},
\end{equation}
because $\omega(X)=p^{\mu}dp_{\mu}=2d(p^2)=0$ on $p^2=m^2$. The 
integral lines of $X$ are 
\begin{equation}
x^\mu(\tau)=p^{\mu}\tau + x_{0}^\mu
\end{equation}
which give the physical motions of the particle.  The space of these
lines is six dimensional (as $p^2=m^2$ and $(p^{\mu},x_{0}^\mu)$
defines the same line as $(p^{\mu}, x_{0}^\mu+p^{\mu}a)$ for any $a$),
and represents the phase space.  The motions are thus the timelike
straight lines in $\cal M$.

Notice that all notions used are completely Lorentz invariant.  A
state is a timelike geodesic; an observable is any Minkowski
coordinate, a correlation is a point in Minkowski space.  The theory
is about correlations between Minkowski coordinates, that is,
observations of the particle at a certain spacetime point.  On the
other hand, the split ${\cal M}={\cal R}\times Q$ necessary to define
the usual hamiltonian formalism, is observer dependent.

\subsubsection{Cosmological model}

Consider a cosmological model in which the sole dynamical variables
are the radius $a$ of a maximally symmetric universe and the spatially
constant value $\phi$ of a scalar field.  Then $\cal C$ has
coordinates $a$ and $\phi$.  The dynamics is given by a single
constraint
\begin{equation}
  H(a,\phi,p_{a},p_{\phi})=0.
\end{equation}
The constraint surface has dimension 3, the phase space has dimension
2, and the motions are curves in the $(a,\phi)$ space.  For each
state, the theory predicts the correlations between $a$ and $\phi$.

\subsection{Two distinct physical meanings of the Lagrangian 
evolution parameter}

Consider the Lagrangian formulation of the dynamics of a pendulum 
with a single Lagrangian variable $\alpha(t)$.  Then the Lagrangian
evolution parameter $t$ is one of the configuration space variables.

On the other hand, consider the Lagrangian formulation of the
cosmological model discussed above, in which the Lagrangian variables
are are $a(t)$ and $\phi(t)$ and the action is reparametrization
invariant.  Or, similarly, consider the reparametrization invariant
Lagrangian for a free relativistic particle 
\begin{equation} 
    S= m \int dt\ \sqrt{(dx^\mu/dt)\ (dx_\mu/dt)} 
    \label{eq:actionparticle}
\end{equation} 
for the four Lagrangian variables $x^\mu(t)$.  In these cases, the
Lagrangian evolution parameter $t$ is unphysical, in the sense that it
has nothing to do with observability.

One should therefore not confuse the $t$ in the first case
with the $t$ in the second case.  They have very different physical
interpretation.  The fact that they are often denoted with the same
letter and with the same name is only a very unfortunate historical
accident.

Now, suppose we are given a time reparametrization invariant
Lagrangian, such as (\ref{eq:actionparticle}) and we blindly Legendre
transform to the relativistic form of hamiltonian mechanics, based the
{\em five dimensional\/} extended configuration space $\tilde{\cal
C}=R\times{\cal M}$ with coordinates $q^a=(t,x^\mu)$.  (Or
$(t,a,\phi)$ for the cosmological model.)  How would we realize that
$\tilde{\cal C}$ has a redundant dimension?  The answer is that we
find two constraints: the constraint (\ref{eq:HH0}) with
$H_{0}=p^\mu p_{\mu}-m^2$, and also the constraint
\begin{equation}
    H_{0}=0. 
\end{equation}
These imply 
\begin{equation}
    p_{t}=0.
    \label{eq:p0=0}
\end{equation}
which, in turn, gives the evolution equation ${dt\over d\tau}=0$ for
the ``time" parameter $t$.  In other words, the Lagrangian parameter
$t$ drops out from the relativistic hamiltonian formalism. 

More precisely, the constraint surface is $\Sigma=R\times{\cal
M}\times {\cal K}_{m}$, but the two-form $\omega$, which is given by
the restriction of $d\theta= dp_{a}\wedge dq^a = dp_{t}\wedge dt +
dp_{\mu}\wedge dx^\mu$ to $\Sigma$, knows nothing about $t$, because
the $dp_{t}\wedge dt$ term vanishes on $\Sigma$.  Therefore, the
variable $t$ decouples completely from the formalism.  This fact
signals that the ``true" extended configuration space $\cal C$ is not
$\tilde{\cal C}=R\times{\cal M}$, but rather ${\cal C}=\cal M$ alone. 
The observables quantities are genuinely independent from the
Lagrangian evolution parameter $t$.  

The partial observables of the relativistic particle are the
quantities $x^\mu$, in any Lorentz frame.  They represent clocks and
devices measuring the spatial position.  The role of $t$ is reduced to
the one of an arbitrary parameter along the orbits and must not be
confused with the role of the time variable $t$ in a nonrelativistic
system, which is a partial observable.  The same happens in general
relativity for the {\em four\/} general relativistic coordinates 
\cite{c2}.

\section{Quantum mechanics}\label{QM}

I conclude with a remark on general covariant quantum mechanics,
although the subject deserves a more complete discussion, which will
be given elsewhere.  Consider a small region $\cal R$ in $\cal C$. 
The region can be taken to represent a certain correlation with a
certain associated experimental resolution.  For a particle $\cal R$
can be a small spacetime region in Minkowski space.  Fix a state in
$\Gamma$.  The motion determines by the state can either intersect
$\cal R$ or not.  Therefore a classical state assigns a yes/no value
to each region $\cal R$ in $\cal C$.  This yes/no value can be seen as
the prediction of whether or not a set of partial observable measuring
devices can give a certain (simultaneous) set of values.  In the case
of the particle, whether or not a detector in the spacetime region
$\cal R$ will detect the particle.

In quantum theory, the predictions of the theory are not
deterministic, but probabilistic.  Accordingly, a quantum state
associates a probability amplitude $P_{\cal R}$ to every small region
of $\cal C$.  How are these probabilities computed?

The quantum dynamics of a system is entirely characterized by a
complex two-point function on the configuration space
\begin{equation}
  W: {\cal C} \times {\cal C} \to C.
\end{equation}
This function is called the \emph{propagator}.  In the classical
limit, the propagator is approximated by the exponential of a solution
of the Hamilton-Jacobi system consider above.
\begin{equation}
  W(x^i,X^i) \sim e^{\frac{i}{\hbar}S(x^i,X^i)}.
\end{equation}
The exact propagator satisfies the Schr\"odinger system 
\begin{equation}
      C\left(x^i,\frac{\partial}{\partial x^i}\right) W(x^i,X^i)=0. 
\end{equation}
The Hilbert space of the theory can be constructed from
$W$ as follows.  Start form a space $\cal E$ of functions $f(x)$ on
$\cal C$ (say smooth compact support).  Consider
the bilinear form on $\cal E$ whose kernel is $W(x,x')$
\begin{equation}
      \langle f |f'\rangle \equiv \int dx \ dx\  \overline{f(x)}\ W(x,x') 
\      f(x'). 
\end{equation}
The Hilbert space $\cal H$ of the theory is obtained by equipping $\cal
E$ with this scalar product, dividing by the zero norm subspace and
completing in norm.  The map 
\begin{eqnarray}
      P: {\cal E} \to {\cal H}\nonumber\\  f \mapsto |f\rangle
\end{eqnarray}
is highly degenerate, and is sometime (improperly) called the
``projector".

To each region $\cal R$ in $\cal C$, we can associate the state
$|{\cal R}\rangle=C_{\cal R}|f_{\cal R}\rangle$, where $f_{\cal R}$ is
the characteristic function of $\cal R$ and $C_{\cal R}=\langle
f_{\cal R}|f_{\cal R}\rangle^{-1/2}$ is the normalization.  For a
region $\cal R$ sufficiently small (smaller than any other quantity in
the problem), the probabilities that express all predictions of
quantum theory are defined by
\begin{equation}
    P_{\cal R} = |\langle {\cal R}|\Psi\rangle|^2. 
\end{equation}
This interpretation postulate reduces to the well known interpretation
of the modulus square of the wave function as spatial probability
density for nonrelativistic systems \cite{rr}.   After the 
correlation determined by $\cal R$ has been verified, the state of 
the system is $|{\cal R}\rangle$. 

In particular, the quantity 
\begin{equation}
    A_{{\cal R},{\cal R}'} =\langle {\cal R}|{\cal R}'\rangle
    \label{eq:main2bis}
\end{equation}
is the probability amplitude to detect the system in the (small)
region $\cal R$ of the extended configuration space, if the system was
previously detected in the (small) region ${\cal R}'$.  This amplitude
can be written explicitly in terms of the propagator 
\begin{eqnarray}
 & &  A_{{\cal R},{\cal R}'} =  
     C_{\cal R}
      C_{{\cal R}'} \int_{\cal R}dx\int_{{\cal R}'} dy \ 
    W(x,y) \nonumber \\ &&\ \ \  = 
      \frac{\int_{\cal R}dx\int_{{\cal R}'} dy \ 
    W(x,y)}{\sqrt{\int_{\cal R}dx\int_{{\cal R}} dy \ 
    W(x,y)}\ \ \sqrt{\int_{{\cal R}'}dx\int_{{\cal R}'} dy \ 
    W(x,y)}}.\nonumber
\end{eqnarray}
See \cite{rr} and references therein.  On covariant approaches to
quantum theory see also \cite{cQM} and references therein.

The important point I want to stress here is that quantum theory
as well can be formulated in a fully covariant language in which time
plays no special role. The fundamental ingredient is once more the
extended configuration space of the partial observables $\cal C$. 
Quantum theory gives probabilities of observing a certain correlation
given that a certain other correlation was been observed.

\section{Conclusion}

The difference between the relativistic definition of state and
observable that we have studied here and the nonrelativistic definition
is the role played by time.  In the nonrelativistic context time is a
primary concept.  Mechanics is defined as the theory of the evolution
in time.  In the definition considered here, on the other hand,there
is no special partial observable singled out as the independent
variable.  Mechanics is defined as the theory of the correlations
between partial observables -- the time variable may be just one among
these.

Historically, the idea that the time independent notion of state is
needed in a relativistic context has been advocated particularly by
Dirac \cite{Dirac2} and by Souriau \cite{Souriau}.  The advantages of
the relativistic notion of state are multifold.  In special relativity
time transforms with other variables, and there is no covariant
definition of instantaneous state.  In a Lorentz invariant field
theory, in particular, the notion of instantaneous state breaks
explicit Lorentz covariance: The instantaneous state is the value of
the field on a simultaneity surface, which is such for a certain
observer only.  The notion of Heisenberg state, on the other hand, is
Lorentz invariant.

This shift in perspective, however, is forced in general relativity,
where the notion of a special spacelike surface over which initial
data are fixed conflicts with diffeomorphism invariance.  A generally
covariant notion of instantaneous state, or a generally covariant
notion of observable at a given time, make very little physical sense. 
Indeed, none of the various notions of time that appear in general
relativity (coordinate time, proper time, clock time) can play the
full role that $t$ plays in nonrelativistic mechanics.  A consistent
definition of state and observable in a generally covariant context
cannot explicitly involve time.

Physically, the reason of this difference is simple.  In
nonrelativistic physics, time and spacial position are defined with
respect to a system of reference bodies and clocks that are always
implicitly assumed to exist and not to interact with the physical
system studied.  In gravitational physics, one discovers that no body
or clock exists which does not interact with the gravitational field:
the gravitational field affects directly the motion and the rate of
any reference body or clock.  Therefore one cannot separate reference
bodies and clocks from the dynamical variables of the system.  General
relativity, or any other general covariant theory, is always a theory
of interacting variables that necessarily include the bodies of
clocks used as references to characterize spacetime points.  In the
example of the pendulum discussed above we can assume that the
pendulum itself and the clock do not interact.  In a general
relativistic context the two always interact, and therefore 
$\cal C$ does not split cleanly into $Q$ and $I\!\! R$.

Mechanics can be seen as the theory of the evolution of the physical
variables in time only in the nonrelativistic limit.  In a fully
relativistic context, mechanics is a theory of correlations between
partial observables, or the theory of the relative evolution of
partial observables with respect to each other.

\end{document}